\documentclass[%
 reprint,
superscriptaddress,
nofootinbib,
 amsmath,amssymb,
 aps,
]{revtex4-2}
\usepackage{todonotes}%
\usepackage{graphicx}
\usepackage{dcolumn}
\usepackage{bm}
\usepackage{amsmath} 
\usepackage{xcolor} 

\newtheorem{theorem}{Theorem}

\newtheorem{corollary}{Corollary}

\newcommand{\Ad}{\mathrm{Ad}}

\newcommand{\Tr}{\mathrm{Tr}}
\newcommand{\su}{\mathsf{su}}

\newcommand{\calN}{\mathcal{N}}
\newcommand{\calW}{\mathcal{W}}

\begin{document}

\preprint{APS/123-QED}

\title{Fundamental Costs of Noise-Robust Quantum Control: Speed Limits and Complexity}

\author{Junkai Zeng}
\email{zengjunkai@iqasz.cn}
 \affiliation{International Quantum Academy, Shenzhen, 518048, China}
\author{Xiu-Hao Deng}
\affiliation{International Quantum Academy, Shenzhen, 518048, China}
\affiliation{Shenzhen Branch, Hefei National Laboratory, Shenzhen, 518048, China}




\date{\today}

\begin{abstract}
Noise is ubiquitous in quantum systems and is a major obstacle for the advancement of quantum information science. Noise-robust quantum control achieves high-fidelity operations by engineering the evolution path so that first-order noise contributions cancel at the final time. Such dynamical error correction typically incurs a time overhead beyond standard quantum speed limits. 
We derive general lower bounds on control complexity that quantify this overhead for quasi-static coherent noise under bounded control amplitude. For a single noise source, we prove a universal time lower bound for first-order robustness and give a constructive scheme that implements any target gate robustly in time \(4T\) plus a constant time. For robustness against an entire noise space, we show dimension lower bounds on the number \(M\) of segments in any mixed-unitary schedule from two mechanism: (i) a \emph{coherent dimension bound} when the error subspace contains an irreducible block isomorphic to \(\su(q)\), and (ii) a \emph{projection dimension bound} when the noise space contains the trace-zero span of orthogonal projectors. Under bounded speed, these bounds on number of segments imply time lower bounds. With only local controls robust against noise space defined on a graph, we obtain a \emph{graph-orthogonality} time bound scales linear with graph chromatic number. We illustrate the bounds through examples. Collectively, these results establish quantitative limitations on the feasibility of first-order noise-resilient operations.
\end{abstract}

\maketitle


\section{Introduction}
Quantum computers promise to perform tasks that are impractical on classical hardware\cite{cheng2023noisy,memon2024quantum}. Yet the fragility of quantum coherence and the ubiquity of noise remain major impediments to building large‑scale processors ~\cite{tripathi2024modeling,burnett2019decoherence,szankowski2017environmental,krantz2019quantum,paladino20141,kumar2016origin}. Noise due to unwanted couplings within the system or to environmental degrees of freedom, fluctuations and drifts of experimental configurations, calibration errors in quantum systems and control hardware, among others, cause deviations from ideal evolution. Although advancement in hardware and manufacturing can reduce these effects ~\cite{huang2024high, wang2024pursuing,kjaergaard2020superconducting}, it is in principle impossible to eliminate noise entirely, as a system perfectly isolated from the environment is also isolated from us, making it impossible to control the system. Quantum error correction, while can actively detect and correct error through error correction code where logic qubit is encoded in an abundant space, requires operations on physical qubit have fidelities above threshold\cite{steane2003overhead,trond2025quantum}.
Quantum error mitigation, on the other hand, aims to reduce the impact of noise without encoding logical qubits or modifying the underlying hardware~\cite{cai2023quantum}. Instead, it relies on post-processing strategies on classical computers, such as extrapolation\cite{giurgica2020digital} probabilistic error cancellation\cite{strikis2021learning}, to reconstruct noiseless expectation values from noisy measurement data. While this approach is hardware-efficient and compatible with near-term devices, it often comes with high classical algorithmic complexity and resource overhead, which limits its scalability for large quantum circuits~\cite{quek2024exponentially}.
Techniques such as dynamical decoupling\cite{khodjasteh2005fault} and dynamically corrected gates (DCGs) \cite{khodjasteh2009dynamically,szankowski2017environmental,Brown2011,kabytayev2014robustness} achieve high-fidelity quantum operations despite the presence of noise with carefully designed control protocols by applying sequences of pulses or continuous drive waveforms, so that the total error accumulated through the evolution reduces to zero when the evolution terminates at the target unitary. 

However, noise-robust control incurs a fundamental overhead: the system must spend additional time to satisfy both the computational objective and the robustness constraints. Classical quantum speed limits (QSLs) bound the minimum time to reach a given unitary, typically connected to time-energy uncertainty relations\cite{pires2016generalized,taddei2013quantum,deffner2017quantum,boscain2021introduction}, but do not incorporate robustness except a few ~\cite{van2018time}. Numerical optimal-control methods (e.g., GRAPE-like algorithms) produce gates by optimizing a cost function that combines gate fidelity and noise suppression\cite{doria2011optimal,khaneja2005optimal,machnes2018tunable,gungordu2022robust,zhang2025smolyak}, but do not reveal intrinsic time limits.
In recent years a new theoretical control framework, termed space curve geometric control (SCGC)\cite{zeng2018general,zeng2019geometric,buterakos2021geometrical,hai2025geometric,walelign2024dynamically,nelson2023designing,dong2021doubly}, was developed where the design of robust control protocol is mapped into the design of a space curve, and the minimal time to achieve robust single-qubit gates was studied with the least action principle to search for the shortest curves that respect system constraints\cite{zeng2018fastest, tang2023designing}. But studies on lower bound on the time required to achieve target unitary while being robust to coherent noise in more general settings remain lacking. 

In this paper, we provide a general framework and explicit bounds for the complexity of first-order robust control against quasi-static coherent noise under amplitude constraints: 
\begin{itemize}
\item For a \emph{single} noise source \(V\), first-order robustness requires time at least \(\pi/u_{\max}\). We also give a constructive robust implementation of any target \(U_T\) with overhead \(\le 4T+\pi/u_{\max}\).
\item For robustness against a \emph{noise space} \(\mathcal N\), we derive segment-count lower bounds that depend on the representation structure of the induced error subspace \(\mathcal W\), and translate them into time bounds using a general speed constraint.
\item Under \emph{local-control} restrictions and pairwise-abelian \(Z\)-type noise on a graph, we prove a graph-orthogonality time bound proportional to the graph chromatic number.
\end{itemize}
These results quantify how controllability, noise-structure, and amplitude bounds jointly constrain the feasibility of fast, noise-robust quantum operations.

This paper is structured as follows.
In Sec.~II we formalize the setup: the control model, noise model, first-order robustness conditions, and the notion of the error subspace.
In Sec.~III we analyze robustness against a \emph{single} noise source, proving a universal time lower bound and presenting a constructive robustification scheme with bounded overhead.
In Sec.~IV we turn to robustness against a general \emph{noise space}, deriving two complementary dimension-based lower bounds---the coherent dimension bound and the projection dimension bound---and mapping schedule complexity to time complexity.
In Sec.~V we study the impact of \emph{local-control constraints}, proving a graph-orthogonality time bound tied to the chromatic number of the noise graph.
Sec.~VI illustrates these bounds with concrete examples. We discuss limitations, point out directions for future research, and conclude in Sec.~VII.


\section{Setup and Assumptions}
We consider a system of \(n\) qubits with Hilbert space \(\mathcal H\cong\mathbb C^d\), \(d=2^n\). The dynamics is governed by a time-dependent Hamiltonian \(H(t)\in i\mathcal K\subseteq \su(d)\), bounded as \(\|H(t)\| \le u_{\max}\),
and generates the propagator \(U(t)\) via \(\dot U(t) = -i H(t) U(t)\). Here $\|\cdot\|$ denotes the operator norm (the largest eigenvalue of the operator)
The reachable group, \(G:=\exp(i\,\mathrm{Lie}(\mathcal K))\subseteq SU(d)\), is the set of unitaries accessible through these controls. 
Typically for a quantum control task, we seek a target unitary $U_T \in G$ at the end of evolution time $T$, while in a broader context we do not assign a specific target unitary, but making use of the expressibility of the reachable group, as in the case of variational quantum algorithms.

\paragraph{Noise model.}
Define the noise space as $\calN \subseteq\su(d)$. We model Quasi-static coherent noise by a traceless operator \(V\in\mathcal N\). During the control, the evolution is perturbed by \(H(t)+\delta V\) with \(|\delta|\ll 1\). The first-order Magnus error (omitting the scalar \(\delta\)) is
\begin{equation}
  E(V;U,T) = \int_0^T U^\dagger(t)VU(t) dt
  =T\mathsf M_T(V), 
\end{equation}
with the averaged adjoint map \(\mathsf M_T:=\frac{1}{T}\int_0^T \Ad_{U(t)}\,dt\).

\paragraph{Robustness conditions.}
First-order robustness against a single \(V\) is \(E(V;U,T)=0\). Robustness against a space \(\mathcal N\) demands \(E(X;U,T)=0\) for all \(X\in\mathcal N\).

\paragraph{Error subspace.}
We define the error subspace as
\begin{equation}
  \calW = \mathrm{span}\{ \mathrm{Ad}_g(V): g\in G, V\in\mathcal N\}\;\subseteq\;\su(d).
\end{equation}
\(\mathcal W\) is \(G\)-invariant. Write its orthogonal decomposition into real irreps as
\begin{equation}
  \mathcal W \;\cong\; \bigoplus_\lambda \big(\mathbb R^{m_\lambda}\otimes W_\lambda\big),
  \qquad \dim W_\lambda = d_\lambda.
\end{equation}
Then each \(W_\lambda\) is also \(G\)-invariant.

\paragraph{Feasibility.}
If \(\mathcal N\) contains a nonzero \(G\)-fixed vector \(A\) (\(\mathrm{Ad}_g(A)=A\) for all \(g\in G\)), then \(\mathsf M_T(A)=A\) and \(A\) cannot be dynamically corrected. In this paper we assume \emph{feasibility}: \(\mathcal N\) has no nonzero \(G\)-fixed vector. This condition holds when we have sufficient large reachable group $G$.

\paragraph{Speed bound.}
For any \(A\in\su(d)\),
\begin{equation}
  \Big\|\frac{d}{dt}\mathrm{Ad}_{U(t)}(A)\Big\|
  \le 2 \|H(t)\| \|A\|
  \le 2 u_{\max}\|A\|.
\label{eq:speed}
\end{equation}
Thus the path $y_V(t)=\Ad_{U(t)}(V)$ is Lipschitz continuous with bounded speed, $\|y_V'(t)\|\le2u_{\max}$. 

\section{Main Result I: Single-Noise Robustness}
We begin our analysis of robustness by considering the simplest case: robustness against a single noise source. Robustness against such a noise source imposes a fundamental lower bound on the evolution time, formalized in the following theorem:
\begin{theorem}[Single-Noise Time Lower Bound]
\label{thm:single-noise}
Let \(V\in\mathcal N\) be a Hermitian involution \(V^2=I\). If robustness condition \(E(V;U,T)=0\) holds, then the evolution time is lower-bounded by \(T \ge \pi/u_{\max}\). 
\end{theorem}

\noindent\emph{Sketch of Proof.}
Let \(y(t):=\Ad_{U(t)}(V)\) with \(\|y(t)\|=\|V\|\). The condition \(\int_0^T y(t)dt=0\) and the speed bound \(\|y'(t)\|\le 2u_{\max}\|V\|\) imply, by the Wirtinger inequality for mean-zero vector-valued functions, 
\(
\int_0^T \|y(t)\|^2 dt \le \frac{T^2}{4\pi^2}\int_0^T \|y'(t)\|^2 dt.
\)
Using \(\|y(t)\|\equiv\|V\|\) and \(\|y'(t)\|\le 2u_{\max}\|V\|\) yields \(T\ge \pi/u_{\max}\). Equality is achieved when the motion of \(y(t)\) is sinusoidal on a great circle in a fixed 2D plane.

The minimal time evolution that satisfies this bound results in an overall identity operation. Implementing a non-trivial target gate \(U_T\neq I\) typically requires a time \(T\) constrained by QSLs, and here in the following theorem we present we introduce a constructive method to implement a robust version of an arbitrary gate.

\begin{theorem}[Robust Gates against Single Noise Source]
\label{thm:4T}
Suppose \(U_T\) can be implemented in time \(T\) under control Hamiltonian \(\|H(t)\|\le u_{\max}\).
Assume there exists a Hermitian involution \(R\) that anticommutes with the noise such that \(R V R = -V\).
Then, a first-order robust realization of \(U_T\) can be implemented in a total time
\(
T^\star \le 4T + \frac{\pi}{u_{\max}}.
\)
\end{theorem}
We proof this theorem by presenting the following explicit control scheme:
\begin{enumerate}
    \item \textbf{Main pass.} Implement \(U_T\) at half speed (time \(2T\)), incurring error \(E_1=2E(V;U_T,T)\).  
    \item \textbf{Flip.} Apply \(R\) via \(e^{-i(\pi/2)R}\) (time \(\pi/2u_{\max}\)), toggling \(V\mapsto -V\).  
    \item \textbf{Palindrome loop.} Run the time-reversed path $L$, whereas $L(t)=u_g(T-t)$ for $0\le t \le T$ and $L(t)=u_g(t-T)$ for $T\le t\le 2T$ (time \(2T\)), producing error \(-E_1\). To implement this, use Hamiltonian $\tilde{H}_L(t)=-RH(T-t)R$ for the first half and $RH(T-t)R$ for the second half.      
    \item \textbf{Flip.} Apply \(R\) again (time \(\pi/2u_{\max}\)) to restore the toggling frame.  
\end{enumerate}
This implementation is feasible in many experiment settings. However, it could be impractical when the control Hamiltonian \(H(t)\) has terms that are fixed or have a fixed sign, since the main pass step requires halving the amplitude, and the Palindrome loop step requires flipping signs of terms that anticommute with \(R\).

The strategy described above is an example of a broader class of methods that append a dynamically corrected identity gate to cancel the error from the main gate operation. Although there may be integrated, more efficient schemes for implementing the robust gate directly, finding them can be a formidable search problem.

To formalize this challenge, consider the concept of \emph{corrector}. A corrector is a Hermitian involution $K$ used to generate an evolution segment for error compensation purposes. The corresponding correction operation generates an error term $F_K=\int_0^s e^{i K\phi(t)}Ve^{-i K \phi(t)}dt$, where $\phi(t)$ is determined by the shape of the control pulse. The set of all possible error vectors $F_K$ generated by corrector \(K\) spans a three dimensional real vector space, $S_K=\mathrm{span}(V, KVK, i[K,V])$, and if $U_T E(V;U,T) U_T^\dagger$ is within $S_K$, the error is correctable by the corrector \(K\). In the case of an $SU(2)$ rotation gate $U_T = e^{-i \Phi T}$ subject to transverse noise \(V\), the choice $K=\Phi$ is sufficient to cancel the error.  A simplest example is $R_X(\phi)$ rotation gate under $Z$-noise (Zeeman-type noise), and the fastest gate implementation that achieves robustness can be found in \cite{zeng2018fastest}. 

In more general scenarios, a single corrector is often insufficient. This is because $E(V;U,T)$ may lie anywhere within a much larger error subspace $\calW$, and the three-dimensional subspace $S_K$ reachable by a single corrector \(K\) often does not contain the error, even though we can freely choose $K$. Consequently, multiple correctors, \(K_1, K_2,...,K_m\), may be necessary. Each additional corrector expands the dimension of the correctable subspace by at most two. Therefore, to cover an error subspace of dimension \(d_\calW\), one needs at least $m\ge (d+1)/2$ correctors. For a known gate protocol $U(t)$ and noise $V$, the form of $E(V;U,T)$ is fixed, and a sequential scheme (see Appendix A) can be implemented to reduce the number of correctors to $\sqrt{d_\calW}$. Due to the fact that $d_\calW$ typically scales exponentially with the number of qubits (worst case $d = 4^N-1$), this is still an exponential scaling. An implication from this scaling is that designing a control scheme perfectly robust against even a single, known noise source often requires solving a search problem in an exponentially large parameter space.

\section{Main Result II: Robustness to a Noise Space}
In this section, we discuss complexity when the goal is to achieve robustness against a noise space \(\calN\), which is determined by the structure of \(\calN\). Generally, theorem \ref{thm:single-noise} still gives a lower bound. When there exists a flip operator within reachable group, \(R\in G\), such that it can flip the entire noise space: \(R X R^\dagger = -X\) for all \(X\in \calN\), then the statement for cancelation of a single noise source still applies. However, this bound does not saturates (even without considering the complexity of implementing a target gate), as such a flipper $R$ does not exist. 

We analyze complexity by simplifying the control protocol to \emph{schedules}. A schedule is a sequence of (well-distinct) unitary operators, \(\{U_k:k=1,...,M\}\), with the sequence length \(M\) measures the schedule complexity. The time-averaged adjoint map can thus be written as a mixed-unitary channel, which is a unital and trace-preserving:
\begin{equation}
  \varepsilon(\cdot) = \sum_{m=1}^{M} p_m\, \Ad_{U_m}(\cdot),
  \qquad p_m>0,\;\sum_m p_m=1,\;\varepsilon(I)=I.
\end{equation}  
The robustness goal is therefore to achieve \(\varepsilon(X)=0\) for all \(X\in\calN\). We analyze \(\varepsilon\) by restricting it to invariant blocks of \(\mathcal W\).

\subsection{Coherent Dimension Bound}
 
\begin{theorem}[Coherent dimension bound]
\label{thm:coherent}
Given a control group \(G\), assume that the generated error space \(\calW \cong \bigoplus_\lambda W_\lambda\) contains a full irreducible block, \(W_\lambda\cong \su(q)\), and is supported on noise space (i.e., \(W_\lambda\subseteq \calN\)), then any schedule obeying robustness against \(\calN\) must have schedule complexity \(M \ge q^2\).
\end{theorem}

\noindent\emph{Sketch of Proof.}
On the irreducible adjoint block, \(\varepsilon\) is a completely positive trace-preserving \emph{mixed-unitary} map that annihilates all traceless elements, making it a depolarizing channel. For any mixed-unitary channel, it has Choi rank \(\le\) the number of unitaries in the mixture ~\cite{girard2022mixed}; on \(\su(q)\) a depolarizing channel requires Choi rank \(q^2\). Hence \(M\ge q^2\). 

This theorem can then be generalized into a no-go statement for robust control against universal noise:
\begin{corollary}[No-go for Full Universal Robustness]
For \(N\) qubits (\(d=2^N\)), if universal robustness is demanded for \emph{all} traceless operators (\(\calN=\su(d)\)) and \(G\) contains \(SU(d)\), then \(M\ge 4^N\). Under a bounded control Hamiltonian, this converts to an exponential time lower bound in \(n\).    
\end{corollary}

\subsection{Projection Dimension Bound}
More often, a full \(\su(q)\) block is not supported on \(\calN\), limiting the application of the coherent dimension bound. We hereby introduce a separate mechanism to derive another dimension bound.
\begin{theorem}[Projection dimension bound]
\label{thm:projection}
Let \(\{\Pi_s\}_{s=1}^{S}\) be pairwise-orthogonal projectors on \(\mathbb C^d\) such that
\[\Pi_s=\Pi^\dagger_s=\Pi_s^2,\quad \Pi_s \Pi_{s'}=0 (s\neq s'),\quad \sum_{s=1}^S \Pi_s=\mathbb I_d\]
Let \(r_s:=\mathrm{rank}\, \Pi_s\) (so \(\sum_s r_s=D\). If the trace-zero span
\(
\mathcal A_0:=\{\sum_{s=1}^{S} c_s \Pi_s:\ \sum_s r_s c_s=0\}
\)
is contained in \(\mathcal N\), then any unital mixed-unitary schedule satisfying \(\varepsilon(X)=0\) for all \(X\in \calN\) must have schedule complexity
\(
M \ge \max_s \lceil d/r_s\rceil.
\)
\end{theorem}

\noindent\emph{Sketch of Proof.}
Each \(\varepsilon(\Pi_s)=\sum_m p_m\, U_m^\dagger \Pi_s U_m\) is positive semidefinite with rank at most \(M r_s\). Robustness against \(\calN\) requires \(\sum_s c_s \varepsilon(\Pi_s)=0\) whenever \(\sum_s r_s c_s=0\). This forces \(\varepsilon(\Pi_s)=r_s Q\) for some fixed \(Q\succeq 0\). Unitality gives \(\sum_s \varepsilon(\Pi_s)=\mathbb I_d\Rightarrow d\,Q=\mathbb I_d\), hence \(\varepsilon(\Pi_s)=\frac{r_s}{d}\mathbb I\) and \(\mathrm{rank}(\varepsilon(\Pi_s))=d\). Thus \(d\le M r_s\) for each \(s\), yielding the claim.

Specifically, if in some $G$-invariant block one can realize \(d_\lambda\) rank-1 projectors with a trace-zero span contained in \(\calN\), then \(M\ge d_\lambda\).

A practical way to find the projection dimension bound is through a spectral decomposition approach: for a noise operator instance $v(a)=\sum_{i=1}^{d_{\mathcal{N}}} a_i V_i \in \mathcal{N}$ where $\{V_i\}_i^{d_\mathcal{N}}$ is a basis of $\calN$, diagonalize it with unitary basis $S$ such that $S v(a) S^\dagger = \tilde{v}(a)=\bigoplus_\lambda \tilde{v}_\lambda(a)$, where each $\tilde{v}_\lambda(a)$ is a diagonal subblock with zero trace and dimension $d_\lambda$ and does not contain any further zero-trace subblock. Within each block one can construct a 1-rank projections by $\{\Pi_{\lambda,s}\} = |s\rangle \langle s|_\lambda$, and the full projection is $S^\dagger (\bigoplus_{\lambda}\{\Pi_{\lambda, s}\})S$. The dimension bound is therefore $M\ge \max_\lambda{d_\lambda}$.

This bound is independent of the control group $G$, as it only uses unitarity and the existence of the commutative block.

\subsection{From schedule complexity to time complexity}
We now convert discrete, schedule-based scheme into continuous dynamics and convert schedule complexity into lower bound on the duration \(T\) under the global amplitude bound \(\|H(t)\|\le u_{max}\). 
For a fixed representative involution noise \(V\in \calN,\, V^2=I\), as in previous sections, set \(y(t):=\Ad_U(t)(V)\), \(\|y(t)\|=1\), so \(y(t)\) is a curve on the sphere-like manifold (the \(G\)-orbit) of the relevant irreducible subblock \(W_\lambda\) of the error space. 
Let \(\mathrm{dist}(y_1, y_2)\) denote the geodesic distance on the \(G\)-orbit, formally \(\mathrm{dist}(y_1, y_2)\ge\|\Theta_{y_1, y_2}\|_\infty\) where \(\|\Theta_{y_1, y_2}\|_\infty\) is the largest principal angle between their +1-eigenspaces. 
Generally, \(\mathrm{dist}(y_1, y_2)\ge\arccos(\langle y_i, y_j\rangle)\) where \(\langle y_i,y_j\rangle=\frac{1}{d_\lambda}\Tr(y_iy_j)\).

To emulate a schedule with complexity number \(M\), partition \([0,T]\) into \(M\) segments \(\{\Delta t_i\}\) and, on each segment, keep \(y(t)\) within a \emph{cap} around a chosen center \(y_i\) corresponding to the schedule step \(y_i=\Ad_{U_i}(V)\), with a symmetric sweep so that \(\int_{\Delta t_i} y(t)dt=w_i y_i\) for some \(w_i>0\). 
When two caps are tangentially touching, the curve can hop from one \(y_i\) to another \(y_j\) continuously, such that the discrete schedule is converted to a continuous dynamics. 
This travel path is lower-bounded by the geodesic distance \(L_{ij}:=\frac{1}{2}(L_i+L_j)\ge \mathrm{dist}(y_i, y_j) \), where \(L_i\) is the length of the symmetric path within the cap \(i\). The curve has total length \(L=\sum_{i=1}^ML_i\ge\sum_{i=1}^M\mathrm{dist}(y_i,y_{i+1})\). Here we adapt a periodic indexing, \(y_{M+1}=y_1\). Combining with the speed bound~\eqref{eq:speed} gives the time lower bound 
\begin{equation}
    T^*\ge \frac{1}{2u_{max}}\sum_{i=1}^M\mathrm{dist}(y_i,y_{i+1})
\end{equation}

Because \(y_i\) in a schedule should be well-separated from each other (so every point contributes to the zero-average), we require\(\langle y_i,y_j\rangle\le0\) for all \(i\neq j\). In this way, \(\mathrm{dist}(y_1, y_2)\ge\arccos(\langle y_i, y_j\rangle)\ge\pi/2\), and the length of the curve \(L\ge M\pi/2\). This gives us a lower bound, \(T^*\ge \frac{M\pi}{4u_{max}}\). 
This bound is generally loose. The tighter bound can be derived by taking into account the structure of the \(G\)-orbit, which depends on both the reachable group \(G\) and noise space \(\calN\). 

As a simple example, consider a noise subspace isomorphic to \(\su(2)\). In this case, the block corresponds to a 2-sphere \(G\)-orbit, and requires a 4-point schedule to dynamically correct. 
One way of making the 4-point schedule a 1-design is to have \(y_i\) form a regular simplex on the 2-sphere (a tetrahedron), making the geometric distance between any two points \(\arccos(-1/3)\), such that $T^*\ge\frac{2}{u_{max}}\arccos(-1/3)\approx3.82/u_{max}$. 
However, this geometrically derived bound is still a loose lower limit. It implicitly assumes the control can steer the noise operator along the geodesics such that each point of \(y_i\) is replaced by a continuous geodesic arc, while a geodesic path for one operator, \(y_1(t)=\Ad_U(t)(V_1)\), is generally not a geodesic path for another operator, \(y_2(t)=\Ad_U(t)(V_1)\). A minimal discrete adjoint 1‑design on \(\su(2)\) is given by the four Pauli unitaries \(\{I,X,Y,Z\}\) (modulo phase), whose continuous implementations include a four \(\pi\)-segments sequence, \(R_X(\pi)\rightarrow R_Z(\pi)\rightarrow R_X(\pi)\rightarrow R_Y(\pi)\). 
That said, geometric 4‑arc paths connecting a regular tetrahedron are useful for the simpler task of dynamically protecting a specific quantum state from \(\su(2)\) noise, as shown in Section VI.

\section{Main Result III: Local Control and Graph-Orthogonality}
The complexity of designing robust controls is further compounded by physical limitations on controllability. In this section, we investigate the consequences of having only local control, where the control group is \(G=\mathrm{SU}(2)^{\otimes N}\) with local bound \(\|H^{(i)}(t)\|\le u_{loc}\) for each qubit \(i\). We consider a common noise model in solid-state qubit systems: local and two-body \(Z\)-type noise whose structure is defined by a graph noise space\(\Gamma=(V,E)\):
\(\calN(\Gamma)=\mathrm{span}\{Z_i\}_{i\in V}\cup \mathrm{span}\{Z_i\otimes Z_j\}_{(i,j)\in E}\). This model accurately represents systems where the residual \(ZZ\)-crosstalk is determined by the physical topology of the chip.

Let \(r_i(t)\) denote the evolution of local noise operator, \(r_i(t)=U^{(i)\dagger}(t) Z_i U^{(i)}(t)\). To efficiently suppress the effect of \(Z_i\)-noise, it is optimal to drive the system with a Hamiltonian that anticommutes with \(Z_i\), such as \(H^{(i)}(t)=\Omega^{i}(t)X_i\). This strategy forces the vector \(r_i(t)\) to rotate on a great circle within the YZ-plane of the Bloch sphere, which is known to minimize the time required for robustness. Therefore, we assume this optimal control form, where \(\Omega^{i}(t) \le u_{loc}\). Under this control, the noise trajectory is confined to \(r_i(t)=\cos(\theta_i(t))Z_i + \sin(\theta_i(t))Y_i\).


\begin{theorem}[Graph-orthogonality bound]
\label{thm:graph}
Let \(\Gamma=(V,E)\) be a finite simple graph representing the noise structure, with chromatic number \(\chi(\Gamma)\). 
If first-order robustness against the graph noise space \(\calN(\Gamma)\) is achieved, and each trajectory \(r_i(t)\) is generated by a local control Hamiltonian with strength bounded by \(u_{loc}\), then the minimum time \(T^*\) for the robust evolution is bounded by:
\begin{equation}
  T^* \ge \frac{\pi}{u_{\mathrm{loc}}}\chi(\Gamma),
\end{equation}
\end{theorem}
\noindent\emph{Sketch of Proof.}
First-order robustness imposes \(\int_0^T r_i(t) dt=0\) for all \(i\) and
\(\int_0^T r_i(t)\otimes r_j(t)\,dt=0\) for all graph edges \((i,j)\in E\). For every edge, these conditions force \(r_i(t)\) and \(r_j(t)\) to be functionally orthogonal, forbidding the two functions be single frequency motions in the same Fourier functional basis, and minimizing \(T\) (hence using the lowest admissible frequency bins) requires assigning distinct frequencies to adjacent vertices. This transforms the problem into one of graph coloring, meaning at least \(\chi(\Gamma)\) distinct Fourier modes \{n\} are needed. The speed cap implies \(\|r'(t)\|\le 2u_{loc}\) constrains the maximum possible frequency. Packing \(\chi(\Gamma)\) orthogonal frequency modes below this speed cap directly imposes a minimum total evolution time, leading to the bound \(T^* \ge \frac{\pi}{u_{\mathrm{loc}}}\chi(\Gamma)\).

We provide a more concrete proof in Appendix B.


\section{Examples}
To illustrate the theory and connect it to concrete situations, we work through several examples. 

\paragraph{Single qubit, \(\calN=\su(2)\).}
The noise space itself forms a coherent bound of \(\su(2)\). Under universal control \(G=SU(2)\), coherent bound gives schedule complexity \(M\ge 4\). A 4-gate dynamical decoupling sequence (XZXY) fulfills this task by using four \(\pi\) rotations, each taking \(\pi/2u_{max}\), for total time \(2\pi/u_{max}\).

If the task is to preserve a given quantum state, e.g., \(|0\rangle\), the 1st-order robust condition becomes \(\langle 0|\int_0^TU^\dagger(t) V U(t)dt |0\rangle\) for \(V \in \calN\), then there exists a faster scheme that saturates the time lower bound with a tetrahedron setting. An explicit construction is to have the curve \(r(t)=(\langle X(t)\rangle,\langle Y(t)\rangle,\langle Z(t)\rangle)\) visit four points, \((x_1,x_2,x_3,x_4)\) sequentially along geodesic lines and forms a loop. This is explicitly shown in Table ~\ref{tab:p}
\begin{table}[]
    \centering
        \begin{tabular}{|c|c|}
        \hline
        Vertices & Coordinate \\
        \hline
         \(x_1\) & \((0,0,1)\) \\
         \(x_2\) & \((\frac{\sqrt{6}}{3},\frac{1}{2\sqrt{3}},-\frac{1}{2})\) \\
         \(x_3\) & \((-\frac{\sqrt{6}}{3},\frac{1}{3},0)\)\\
         \(x_4\) & \((0, -\frac{\sqrt{3}}{2},-\frac{1}{2})\)\\
        \hline
        \end{tabular}     
    \caption{Vertices along the curve. When arrived at a vertex, the curve make a sharp turn and proceed to the next vertex along geodesic line.}
    \label{tab:p}
\end{table}
\begin{table}[]
    \centering
    \begin{tabular}{|c|c|}
    \hline
    \(H(t)\) segment & Form \\
     \hline
    \(K_1\) & \(\frac{1}{3}X-\frac{2\sqrt{2}}{3}Y\) \\
    \(K_2\) & \(-\frac{1}{3}X-\frac{\sqrt{2}}{3}Y-\frac{\sqrt{6}}{3}Z\)\\
    \(K_3\) & \(\frac{1}{3}X+\frac{\sqrt{2}}{3}Y-\frac{\sqrt{6}}{3}Z\) \\
    \(K_4\) & \(X\) \\
     \hline
\end{tabular}
    \caption{Drive direction of Hamiltonian that saturates time lower bound for state preserving. Within time segment \(i\), the system is driven through control Hamiltonian \(H(t)=u_{max}K_i\).}
    \label{tab:ham}
\end{table}
\begin{figure}
    \centering
    \includegraphics[width=0.9\linewidth]{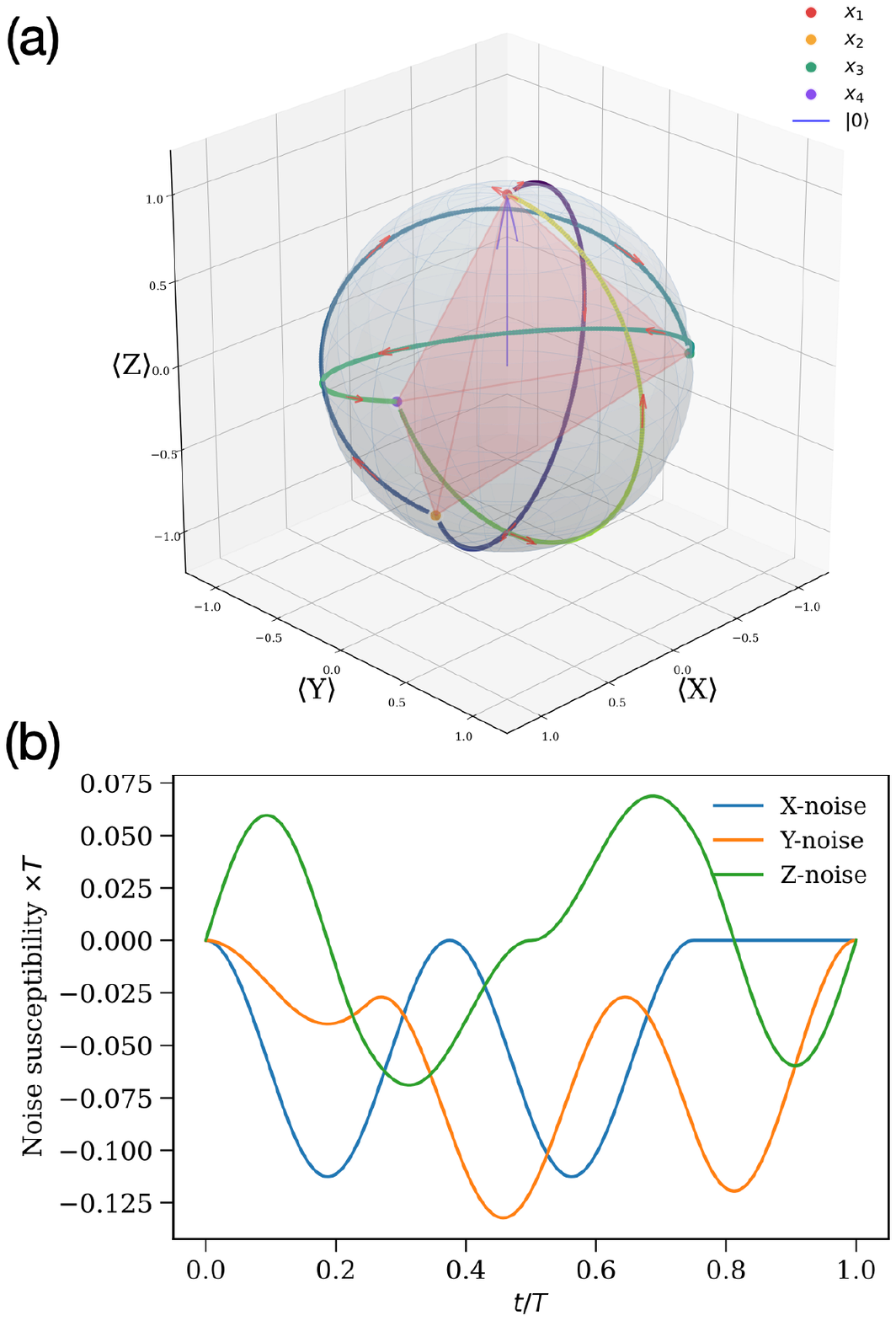}
    \caption{Tetrahedral state‑preservation trajectory and noise susceptibility. (a) shows Bloch-sphere trajectory of \(|0\rangle\) (north pole) under the tetrahedral control. The path follows four great-circle arcs that sequentially pass through tetrahedron vertices \(\{x_1, x_2,x_3,x_4\}\). (b) visualize the error accumulated through the dynamics, \(E_0(t;V)=\langle0|\int_0^t U^\dagger(\tau)V U(\tau)d\tau|0\rangle\) for \(V\in\{X,Y,Z\}\). All three plots vanishes at the end-time, illustrating the protection against \(\su(2)\) noise space for state \(|0\rangle\).} 
    \label{fig:qubitpath}
\end{figure}
The total time is \(T^*=2\arccos(-1/3)/u_{max}\). The control Hamiltonian is to drive along four \(K_i\)'s sequentially, explicitly shown in Table~\ref{tab:ham}, and each drive lasts for \(T^*/4\). We also explicitly visualize this qubit dynamics in a Bloch sphere as well as the accumulated error (noise susceptibility) in Fig. ~\ref{fig:qubitpath}.

\paragraph{Two qubits with Ising noise.}
Let \(\mathcal N=\mathrm{span}\{Z_1,Z_2,Z_1Z_2\}\). The projection bound with rank-1 projectors gives \(M=4\), achieved by \(\{U_i\}=\{I, X_1, X_2,X_1X_2\}\), implemented with alternating gates \(X_1\) and \(X_2\), each taking time \(\ge\pi/2u_{max}\), making the total time \(T^*\ge 2\pi/u_{max}\). This lower bound on time is tight. Through the graph-orthogonality bound, theorem~\ref{thm:graph} with \(\chi(\Gamma)=2\) gives \(T^*\ge 2\pi/u_{loc}\); with \(u_{loc}=u_{max}\), this also yields \(2\pi/u_{max}\).

\paragraph{1-D qubit chain with local control.}
With local and nearest-neighbor \(ZZ\)-residual coupling noise, a qubit chain forms a graph \(\chi(\Gamma)=2\). Assigning identical local paths on even sites and a second path on odd sites achieves robustness with minimal time \(T^*\ge 2\pi/u_{loc}\), matching the two-qubit case. When there is an odd number of qubits and the chain forms a closed ring (the first qubit has residual \(ZZ\)-coupling with the last qubit), the minimal time increases to \(T^*\ge 3\pi/u_{\mathrm{loc}}\) as the chromatic number is now 3.

\paragraph{Complete graph and dense couplings}
Consider $N$ qubits with all-to-all pairwise $ZZ$ residual couplings and local controls. This forms a complete graph so $\chi(\Gamma)=N$. Theorem ~\ref{thm:graph} then gives \(T^*\ge N/u_{loc}\).

\section{Discussion}
Our bounds quantify intrinsic overheads for first-order robust control and clarify how they scale with noise-space structure, controllability, and locality. They (i) recover known two- and four-step toggling times in simple cases, (ii) identify exponential obstacles for worst-case noise spaces, and (iii) expose graph-theoretic obstructions under local control.
We point out that, when other realistic factors are considered, e.g., finite control bandwidth, time complexity could be further impacted. Second, while we provide a schedule complexity number, a concrete recipe to build such a schedule is not discussed in detail. When mapping from schedule to continuous dynamics, one needs to consider the geodesic distance within an error space between different points\cite{nielsen2005geometric,dowling2006geometry}, which is a nontrivial task and affects the strategy on designing the optimal schedule (and the corresponding optimal control scheme). Finally, the bound we provide saturates mostly when the dynamics form an identity (trivial dynamical decoupling). Although we pointed out, searching for robust control scheme for a target gate, even against a fixed noise source is hard, this hardness can be potentially reduced by 1) instead of chasing perfect robustness, allowing a finite noise susceptibility such that \(\|\mathsf{M}_T(V)\|\le\epsilon\); 2) instead of searching control protocol generated through corrector \(e^{-i K \phi(t)}\), search within parameterized unitary space (the Palindrome loop we introduced is one such example). Through the geometric control framework, dynamics that robust against a noise operator always correspond to a closed curve within an error space, and this can potentially be used as a source to search for more robust gate implementation.

\appendix
\section{Cancelling a given first–order error with at most \(d+1\) correctors}
\label{app:linear-in-d}

We give a constructive scheme and a clean bound: for any \emph{fixed} first–order error
\(E \in \su(d)\) (with \(d=2^n\)), one can cancel \(E\) using at most \(d+1\)
distinct correction axes \(\{K_j\}\) (i.e., at most \(d+1\) segments of the form
\(e^{-i K_j t}\)). In many instances, far fewer axes suffice. Thus the \emph{per–instance}
complexity scales linearly in \(d\), not in \(d^2\).

\paragraph{Setup and notation.}
Fix a Hermitian involution \(V\) with \(V^2=I\) (e.g., a Pauli string).
Let \(\mathcal H_+\oplus \mathcal H_-\) denote the +1 and -1 eigenspaces of \(V\)
(\(\dim\mathcal H_+=\dim\mathcal H_- = d/2\)). In this \(V\)–eigenbasis, write the
known first–order error as
\begin{equation}
    E =
\begin{bmatrix}
A & B \\
B^\dagger & D
\end{bmatrix},
\qquad \Tr(A)+\Tr(D)=0.
\end{equation}
We will construct a finite sequence of Hermitian involutions \(K_j\) and dwell waveforms
so that the net first–order correction equals \(-E\).

\paragraph{One axis spans a 3‑dimensional knob space.}
For any Hermitian involution \(K\) (written in the same block split as
\(K=\left[\begin{smallmatrix} X & Y\\ Y^\dagger & Z \end{smallmatrix}\right]\)),
the integral of the toggled operator along \(K\) satisfies
\begin{equation}
    \int_0^{s} e^{+iK\phi(t)} V e^{-iK\phi(t)} dt
=
\alpha V + \beta K V K + \gamma i[K,V],
\end{equation}
for suitable real \(\alpha,\beta,\gamma\) determined by the waveform \(\phi(t)\).
In the \(V\)–basis, these three generators have the block forms
$V=\begin{bmatrix} I & 0\\ 0 & -I\end{bmatrix}$, $
i[K,V]=\begin{bmatrix} 0 & -2i Y\\ 2i Y^\dagger & 0\end{bmatrix}$, $
K V K=
\begin{bmatrix}
I-2YY^\dagger & \star\\
\star^\dagger & -I+2Y^\dagger Y
\end{bmatrix}$,
with \(X=\sqrt{I-YY^\dagger}\), \(Z=-\sqrt{I-Y^\dagger Y}\), and
\(\sqrt{I-YY^\dagger} Y=Y\sqrt{I-Y^\dagger Y}\) (the standard Halmos/CS
construction). Conversely, any contraction \(Y\) (\(\|Y\|\le 1\)) produces a valid
Hermitian involution \(K\) via these formulas.

The construction proceeds in two steps: (i) cancel the off–diagonal block \(B\) with one
axis; (ii) match the diagonals with at most \(d-1\) additional axes \(\{K_j\}\), chosen
so they do \emph{not} re‑introduce off–diagonals; finally (iii) adjust a scalar baseline
along \(V\).
This yields a total of at most \(1+(d-1)+1 = d+1\) axes.

\begin{theorem}[At most \(d+1\) correctors]
\label{thm:d-plus-one}
Let \(V^2=I\) and \(E=\left[\begin{smallmatrix}A&B\\ B^\dagger&D\end{smallmatrix}\right]\)
be given in the \(V\)–basis. There exist Hermitian involutions
\(K_1,\dots,K_m\) with \(m\le d+1\) and real coefficients
\(\alpha,\beta_2,\dots,\beta_m,\gamma_1\) such that
\begin{equation}
\alpha V + \gamma_1  i[K_1,V]+ \sum_{j=2}^{m}\beta_j K_j V K_j =E.
\end{equation}
Consequently, applying the corresponding correction segments (with these weights) cancels
the given error to first order.
\end{theorem}

\emph{Step 1 (off–diagonals in one shot).}
Choose \(\gamma_1\neq 0\) and set \(Y_1:=\frac{i}{2\gamma_1} B\).
For \(|\gamma_1|\) large enough, \(\|Y_1\|\le 1\), so the Halmos construction yields a
valid involution \(K_1\) with that off–diagonal. Then
\begin{equation}
    \gamma_1 i[K_1,V] =  
\begin{bmatrix} 0 & B\\ B^\dagger & 0\end{bmatrix},
\end{equation}
which exactly matches the off–diagonal block of \(E\).

\emph{Step 2 (diagonals with at most \(d-1\) axes and no new off–diagonals).}
Pick a scalar \(s>0\) (e.g. \(s\ge \|E\|\)) and define
\begin{equation}
A' := \frac{1}{2}(sI-A)\succeq 0,\qquad
D' := \frac{1}{2}(sI-D)\succeq 0.
\end{equation}
Because \(\Tr(A)+\Tr(D)=0\) and \(\dim\mathcal H_+=\dim\mathcal H_-=d/2\),
we have \(\Tr(A')=\Tr(D')=:S\).
There exists a “shared–weights” rank–one decomposition with at most \(d{-}1\) terms:
\begin{equation}
A'=\sum_{j=2}^{m} w_j\,u_j u_j^\dagger,\qquad
D'=\sum_{j=2}^{m} w_j\,v_j v_j^\dagger,
\end{equation}
with \(w_j>0\), \(\sum_{j=2}^{m} w_j=S\), and \(m\le d\).
obtained by taking the common refinement of the spectral cumulative sums of \(A'\)
and \(D'\) on \(\mathcal H_+\) and \(\mathcal H_-\).

For each \(j\ge 2\), set $Y_j := u_j v_j^\dagger\qquad(\|Y_j\|=1)$, 
and build \(K_j\) by the Halmos/CS formulas.
Because \(Y_j\) is a rank–one \emph{partial isometry}, we have
\(\sqrt{I-YY^\dagger} Y = 0\) and \(Y \sqrt{I-Y^\dagger Y}=0\), hence the
\emph{off–diagonal blocks of \(K_j V K_j\) vanish}. Its diagonal blocks are
\begin{equation}
(K_j V K_j)_{++}=I-2 u_j u_j^\dagger,\qquad
(K_j V K_j)_{--}=-I+2 v_j v_j^\dagger.
\end{equation}
Therefore, choosing \(\beta_j:=w_j\) makes the diagonal contribution of
\(\beta_j K_j V K_j\) equal to
\(
\begin{bmatrix}
-2 w_j u_j u_j^\dagger & 0\\
0 & 2 w_j v_j v_j^\dagger
\end{bmatrix}.
\)
Summing over \(j=2,\dots,m\) gives the desired pair \((-2A',2D')\).

\emph{Step 3 (baseline along \(V\)).}
Finally, set \(\alpha:=s-\sum_{j=2}^{m}\beta_j\) so that the total diagonal is
\begin{equation}
\alpha V+\sum_{j=2}^{m}\beta_j K_j V K_j=
\begin{bmatrix}
\alpha I -2A' & 0\\ 0 & -\alpha I + 2D'
\end{bmatrix}=
\begin{bmatrix} A & 0\\ 0 & D\end{bmatrix}.
\end{equation}
Together with Step~1 this reproduces the full \(E\).
Counting axes: one for Step~1, at most \(d-1\) for Step~2, and one for the baseline
\(\alpha V\), for a total of \(m\le d+1\).

\paragraph{Remarks.}
(i) If \(B=0\), the off–diagonal step is unnecessary.
(ii) In many instances one can choose \(s\) so that \(\alpha=0\), reducing the count to
\(m\le d\).
(iii) The construction is \emph{instance–wise}: it uses at most \(d+1\) axes for the
\emph{given} \(E\). It does not attempt to span all of \(\su(d)\) a priori.

\section{Proof of Graph-orthogonality bound}
In this section, we provide proof on the linear time scaling of robust control time complexity about the graph chromatic number, for system with pairwise $ZZ$ residual couplings and local controls only. 

We use a preliminary from \emph{high-dimension Poincare inequality}. Let \(H_0\) be a Hilbert space with orthonormal basis: \(\phi_{n,1}=\sqrt{\frac{2}{T}}\cos(n\omega t)\), \(\phi_{n,2}=\sqrt{\frac{2}{T}}\sin(n\omega t)\) with \(n>1\). This way all \(f\in H_0\) is mean-zero (\(\int f=0\)). For Laplacian \(-\frac{d^2}{dt^2}\), \(H_0\) has discrete spectrum   \(\lambda_1\le \lambda_2\le \dots\). Let \(U\subset H_0\) be any subspace with \(\dim U=m\), for every function \(f\in H_0\) with \(f\perp U\), 
\begin{equation}
    \int_0^T f(t)^2 dt\le \frac{1}{\lambda_{m+1}}\int_0^T |f'(t)|^2 dt
\end{equation}
Because \(\lambda_1=\lambda_2=\omega^2,\, \lambda_3=\lambda_4=(2\omega)^2,\dots\), the explicit bound is
\begin{equation}
\label{eq:poincare}
    \int_0^T f(t)^2dt\le \frac{T^2}{4\pi^2 k^2}\int_0^T |f'(t)|^2 dt,\qquad k=\lfloor \frac{m}{2}\rfloor+1,
\end{equation}

For mean-zero \(r_i=(x_i,y_i)\), robustness against two-body noise requires \(\int_0^T r_i\otimes r_j=0\) with \(|r'(t)|\le B\) (\(B=2u_{loc}\) in our set up). This way, every element in \(r_i\) is \(L^2\)-orthogonal to every element of \(r_j\). Consider a clique on the graph \(\Gamma\) (all-to-all sub graph) of size \(N\), this makes all \(r_i\)'s mutually orthogonal to each other. Then each element of \(r_k\) is orthogonal to \(U_{k-1}:=\mathrm{span}\{x_i,y_i:1\le i\le k-1\}\subset H_0\), with \(\dim U_{k-1}=2(k-1)\). Eq.~\ref{eq:poincare} (with \(m=2(k-1)\) hence gives 
\begin{equation}
    \int_0^T |r_k|^2dt \le \frac{T^2}{4\pi^2 k^2}\int_0^T |r'(t)|^2dt.
\end{equation}
Using \(|r_k|=1\), \(\int_0^T |r_k|^2dt=T\), this becomes
\begin{equation}
T\le \frac{T^2}{4\pi^2 k^2}\int_0^T |r'(t)|^2dt \le\frac{B^2T^3}{4\pi^2 k^2}.
\end{equation}
yielding a lower time limit \(T\ge \frac{2\pi k}{B}\). The bound saturates by pure rotations, \(r_k(t)=(\cos(k \omega t), \sin(k \omega t))\), \(\omega=1/N\), \(T=2\pi N/B\).

For most (realistic) cases, the clique number equals chromatic number. To prove the scaling in chromatic numbers, use pure rotation on each \(r_i\), and each distinct frequency number \(n\) corresponds to a color. Orthogonality implies that two adjacent vertices cannot have the same frequency number, hence proving the scaling on the chromatic number.

\acknowledgments{This work was supported by the National Natural Science Foundation of China (Grants No. 12404566).}

\bibliography{apssamp}

\end{document}